# TESTING OF THE 2.6 GHz SRF CAVITY TUNER FOR THE DARK PHOTON EXPERIMENT AT 2 K

C. Contreras-Martinez[†], B. Giaccone, I. Gonin, T. Khabiboulline, O. Melnychuk, Y. Pischalnikov, S. Posen, O. Pronitchev, and JC. Yun, Fermilab, Batavia, Illinois, USA

## Abstract

At FNAL two single cell 2.6 GHz SRF cavities are being used to search for dark photons, the experiment can be conducted at 2 K or in a dilution refrigerator. Precise frequency tuning is required for these two cavities so they can be matched in frequency. A cooling capacity constraint on the dilution refrigerator only allows piezo actuators to be part of the design of the 2.6 GHz cavity tuner. The tuner is equipped with three encapsulated piezos that deliver long and short-range frequency tuning. Modifications were implemented on the first tuner design due to the low forces on the piezos caused by the cavity. Three brass rods with Belleville washers were added to the design to increase the overall force on the piezos. The testing results at 2 K are presented with the original design tuner and with the modification.

## INTRODUCTION

A dark photon is a hypothetical particle that weakly couples to ordinary matter and is an extension of the standard model (SM) of particle physics [1]. At FNAL the search for dark photons consists of two high $Q_o$ single cell 2.6 GHz SRF cavities, one of which will be called the emitter cavity and the other the receiver cavity. This search is one of several which are known as "light-shining-through-wall" experiments [2]. The emitter will be powered on creating an electromagnetic field ($TM_{010}$) inside the cavity, the field from this emitter cavity (005) acts as a source of dark photons that can be emitted outside the cavity. The dark photons can penetrate the receiver (006) cavity and be converted to SM photons [2] at the same frequency as the emitter cavity. The receiver cavity thus acts as the sensor. The search for dark photons will take place in the vertical test stand (VTS) at 2 K and inside a dilution refrigerator.

Precise frequency tuning is required for these two cavities so they can be matched in frequency. The 2.6 GHz tuner consists of two titanium (Ti) brackets that attach to the cavity via studs. A cooling capacity constraint on the dilution refrigerator only allows piezo actuators to be part of the tuner system. Three long piezo capsules are used for resonance control. The capsules each contain two 10 × 10 × 36 mm PICMA stacks glued together [3]. The nominal displacement is 38 µm per stack when applying a voltage of 0 to 120 V at room temperature and the total stroke is 76 µm per capsule. The piezo capsules are placed in between the two Ti brackets via brass screws that have a ceramic ball (See Fig. 2a). The piezo capsules stretch the cavity increasing the frequency when applying a positive voltage. At temperatures below 77 K, the voltage can go down to -120 V. Applying a negative voltage to the piezo shrinks it with proper piezo preloading, then in this regime the cavity frequency can also be decreased. A schematic of the piezo with different voltages applied to it is shown in Fig. 1.

Before the cavities are placed inside the Dewar for testing the piezos are preloaded by using the screws on the Ti bracket. Since the piezos are stiffer than the cavity and Ti bracket the cavity stretches when the screws are adjusted, the cavity then applies a force on the piezos. The purpose of the preload is to overcome the initial deformation of the Ti bracket and avoid gaps due thermal shrinkage of the different materials once cooled to 2 K. The piezos must have a preload of 1 kN (based on the experimental results from this paper) each to avoid these issues. The preload force applied to the piezos is determined by the 35 kN/mm cavity stiffness and the 10 kHz/µm longitudinal frequency deformation sensitivity [4]. Once the proper preload has been set the cavity-tuner system has a piezo stroke to cavity deformation efficiency of 70 - 80 % [4]. An estimate of the tuner range can be calculated based on the stroke of the piezo. At 2 K the piezo stroke is 20 % from the room temperature value resulting in a stroke of 15.2 µm from 0 to 120 V for the capsule. The efficiency of the tuner is 70 % thus the total range expected is 212 kHz $\pm$30 kHz (from -120 V to 120 V).

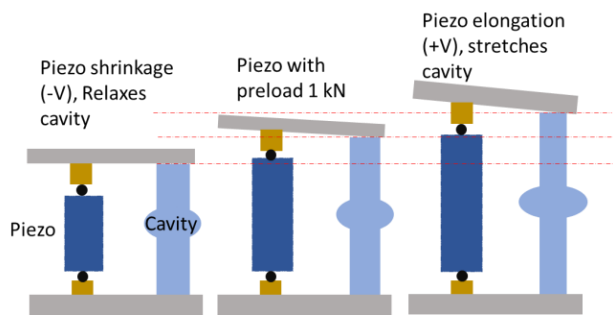

Figure 1: Schematic of piezo with 1 kN preload. The piezo elongation under a positive voltage will cause the cavity to stretch. A negative voltage will relax the cavity.

## INITIAL DESIGN

Testing at 2 K requires that the piezo preload set at room temperature is preserved. The different shrink rates of the materials of the tuner can cause the preload to decrease at 2 K. This issue is avoided by using a stepper motor such as

This material is based upon work supported by the U.S. Department of Energy, Office of Science, National Quantum Information Science Research Centers, Superconducting Quantum Materials and Systems Center (SQMS) under contract number DE-AC02-07CH11359 in the 1.3 GHz dark photon search [5]. In this case, the motor can be used to compress the piezos further if needed. Due to the cryogenic constraint of the dilution refrigerator, only piezos can be used in this setup. The first test at 2 K resulted in a piezo stroke of ~ 10 kHz. The preload on the three piezos was 1.05 kN (350 N per piezo), and a frequency match between the two cavities was not possible. Since brass shrinks twice as much compared to niobium from 293 K to 4 K [6] the initial preload done at room temperature is lost. The pressure outside the cavity changes from room temperature to 2 K, this change also caused the preload to decrease.

[†] ccontrer@fnal.gov

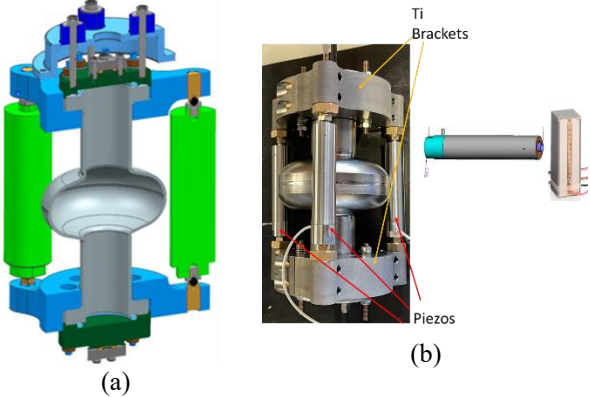

Figure 2: (a) Cross-section view of the cavity. (b) Picture of the cavity with piezos installed. The picture of the actuator with a single stack is also shown.

During the second test at 2 K, the piezo preload at room temperature was 2.45 kN (816 N per piezo), again this was done by stretching the cavity. In this case, both cavities were able to be matched to the same frequency (see Fig. 3). Note that not all the piezo stroke went directly to stretch the cavity as shown by the different frequency trends in Fig 3. The figure shows three trends for both cavities 005 and 006. From -50 V to 72 V (-120 V to 72 V for cavity 006), the figure shows that piezos are not fully engaged. From 72 V to 120 V, the piezos are fully engaged and the piezo sensitivity is 970 Hz/V. For cavity 005 when applying a piezo voltage below -50 V the cavity frequency is constant indicating that all the piezos have small to no preload with the cavity. For cavity 006 below 72 V, the cavity displays two frequency trends, the different trends could be due to the number of piezos engaged. For example, only 2 of the three piezos could be engaged. Note that although the cavity frequencies can be matched only the region with full engagement can be used. When the piezos are fully engaged the cavity is less susceptible to microphonics and df/dp since the piezos provide large stiffness.

The culprits for not achieving full piezo engagement are the same as in the first test. When the cavity was stretched by 70 μm to achieve a force of 2.45 kN (816 N per piezo) the inelastic limit was reached. Thus a 70 μm stretch is then equivalent to a stretch of ~ 40 μm since the cavity does not go back to the original configuration. The force is then less than what was initially calculated. This inelastic deformation of the cavities also complicates the cavity frequency match at room temperature since each large piezo preload iteration could result in a large gap between the frequencies. Using the large preload method will not be able to provide enough force to overcome the initial deformation of the Ti bracket and thus improve the efficiency of the piezo stroke.

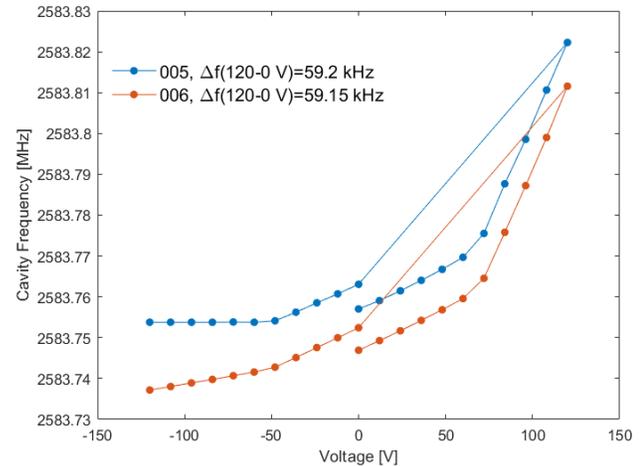

Figure 3: Results of both cavities when the piezo preload was 2.45 kN. Note that three different trend lines are observed.

## TUNER MODIFICATIONS

To increase the total preload force on the piezos the preload force caused by the cavity was complemented with Belleville washers. The Belleville washers act as a spring and when compressed provide a force that acts on the piezos. Since the Belleville washers provide a large force there is no need to stretch the cavity to the inelastic regime. The tuner modification design used long brass rods that hold the Belleville washers to compress the piezos. The rods were held in place via an aluminum piece that is connected to the Ti bracket as shown in Fig. 4. The stiffness of this aluminum piece is 19.2 kN/mm based on the simulation. Most of the deformation goes to the location of the piezo and not the cavity. With this modification, the piezos will be compressed with the force of the cavity and the force of the Belleville washers.

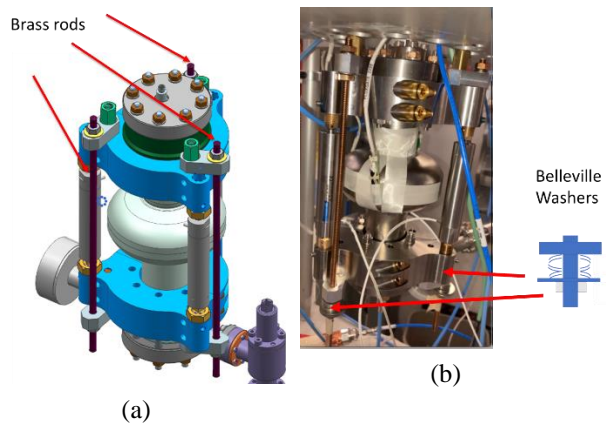

Figure 4: (a) Second design of cavity tuner with brass rods. (b) Picture of cavity installed on the VTS insert, schematic of how the Belleville washers go on the brass rod.

The piezos are first preloaded as in previous iterations, first by stretching the cavity. The cavity is stretched with the piezos by 30 μm which is equivalent to a total force of 1.05 kN. The Belleville washers were arranged to provide a stiffness of 2.7 kN/mm per rod. To provide a force on the piezo one of the nuts on the rods is locked while the other nut on the opposite side is used to squeeze the washers thus providing a force (see Fig. 4b). The amount of force applied was calculated using the pitch of the nut to get the displacement. Additionally, the cavity frequency was monitored to verify the displacements calculated were correct. A schematic of the force from the Belleville washers and cavity is shown in Fig. 5.

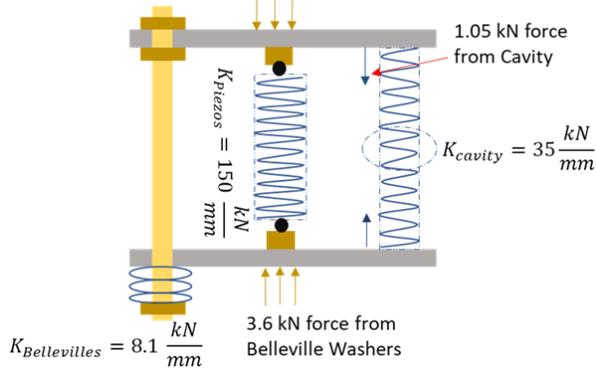

Figure 5: Schematic of the Belleville washers, piezos, and cavity equivalent stiffness.

The force on the piezos from the Belleville washers was set to 3.6 kN at room temperature (1.2 kN each). The total force on the piezos from the cavity and the Belleville washers is then 4.65 kN (1.55 kN per piezo). Note that these are only the forces at 293 K. At 2 K the forces are different. When the cavity is tested at 2 K the pressure outside the cavity (the inside of the cavity is under vacuum) changes to 23 Torr. This change in pressure from 1 bar to 23 Torr causes the cavity to elongate. This elongation will compress the Belleville washers thus adding more force, but at the same time the piezos will lose some contact with the cavity. The brass rods will also shrink more than the cavity, the calculated shrinkage is roughly 400 μm. The Belleville washers will be compressed by this amount and cause a force of 3.25 kN. The total force on the piezos is then at most 7.9 kN (2.33 kN per piezo). Note that a large force on the piezo does not affect the piezo stroke as demonstrated in [7].

The result of the tuner modifications is shown in Fig. 6, note that the piezo range is larger than that of the original design with a 2.45 kN preload with just the cavity providing a force. The hysteresis curve for 005 is different from 006 between 0 to 100 due to the long wait time between each data point. The total frequency range is now 176 kHz for cavity 006 and 158 kHz for 005 compared to the 60 kHz on the original tuner design. The difference in range between the cavities is due to the longer wait times between the data points. This frequency range is within the error (212 kHz ±30 kHz) from the calculated range based on the piezo stroke discussed in the introduction. The piezo can now produce a cavity frequency shift when applying a negative voltage. This design modification allowed both cavities to be matched in frequency. The differences between both tuner designs are summarized in the table below.

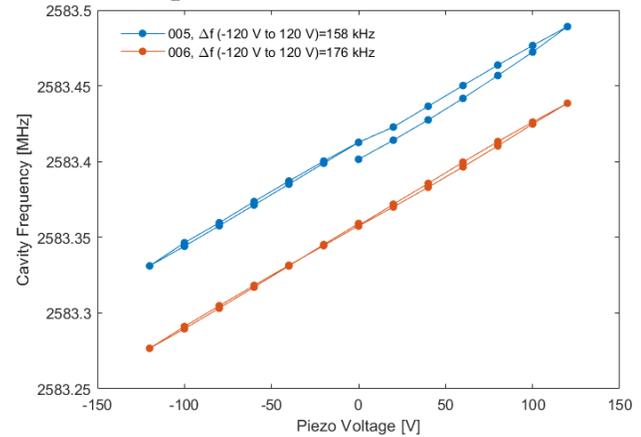

Figure 6: Results of frequency change with piezo with the second design. The different curves are due to long wait times to get the data for the 005 cavity.

Table 1: Comparison of the total forces on the piezo at 293 K and piezo sensitivity when piezos are fully engaged with a voltage range of -120 V to 120 V.

|  | Total Force on Piezos [kN] | Piezo Sens. [Hz/V] | Full Range [kHz] |
| --- | --- | --- | --- |
| 1st | 2.45 | 970 | 60 |
| 2nd | 4.65 | 673 | 176 and 158 |

## CONCLUSION

The 2.6 GHz cavities that will be used for dark photon search were tested at 2 K. Initial tests showed that the tuner design needed improvements so that the piezos would have full engagement. Both cavities were matched with the same frequency after tuner modifications were implemented. The brass rods and Belleville washer design of the tuner yielded a range of 176 kHz. With this modification, a smaller cavity stretch is now required which prevents an inelastic deformation.